\documentclass{ws-procs975x65}

\begin{document}
\title{NON-BPS BLACK HOLES AND SELF-INTERACTING FUNDAMENTAL STRINGS}
\author{DIEGO VALERIO CHIALVA} 
\address{Nordita Institute, Roslagstullsbaken 23, 10691 Stockholm,
  Sweden. E-mail: chialva@nordita.org .}

\begin{abstract}
The string-black hole
correspondence principle can be investigated in the non-BPS scenario
by studying the string configuration and entropy when the string
coupling is slowly 
increased. Through a
rigorous analysis, it is shown how
an ensemble of string states at fixed
mass and Neveu-Schwarz charges gets dominated in any
dimension by compact states for which the one-loop corrections are
important (possibly signaling the transition to a black hole
regime/description)
and with a size (spread) within the horizon radius of
the expected correspondent black holes.
\end{abstract}

\bodymatter
\bigskip

The relation between the 
Hawking-Bekenstein entropy of black holes and the counting of
microstates is
one of the most studied topics in black holes physics. In this short
communication we deal with the non-BPS  
scenario in the setup of closed superstring theory. In doing so, we
touch upon some important questions:
how to individuate the 
relevant microstates, how and when the two descriptions (general relativity
and quantum string) match and, finally, how geometric
features of black 
holes arise from the quantum description.

Within string
theory, the string-black hole correspondence principle
\cite{correspondenceprinciple} states that a
classical black hole is in one-to-one correspondence with an ensemble
of string and/or D-brane states (depending on its type of charges).

Two quantities in particular play a key role in the analysis: the
string coupling $g_s$\footnote{Linked 
to the Newton's constant ad the string length at the perturbative
level as $G_N\sim g_s^2 (\alpha')^{\frac{d-1}{2}}$ in $d=D-1$ extended
spatial dimensions.} 
and the black hole horizon radius $R_{BH}$, especially in relation
with the string length scale\footnote{In what follows
the horizon radius is to be intended in the {\em string frame}.
The general relativity description is valid 
when the curvature at $R_{BH}$ (in the string frame) is 
smaller than the string scale, this yields the condition $R_{BH} > l_s$.} $l_s
=\sqrt{\alpha'}$. 
  
The string and the black hole entropy are conjectured to become equal
at a correspondence point $g_s \sim 
(M^2-Q^2)^{-\frac{1}{4}} \equiv g_c$ in parameter space, where $M$ is
the mass of the string states in the ensemble, equal to the black
hole's one at this  
point, and $Q^i$ are the Neveu-Schwarz
charges\footnote{Here, for simplicity 
$|Q^i_L|=|Q^i_R|=|Q^i|, \, Q^2= \sum (Q^i)^2$, where $L(R)$ refer to the
  (anti)holomorphic 
part of the string.}. Also, at 
this point $R_{BH}$ approaches $l_s$.

There are major obstacles to verifying such a statement within string theory. 
We must show what happens to the relevant ensemble of string
states and their entropy\footnote{We will consider
  single-string entropy, which dominates the contribution to the total
  entropy.} when the string coupling (adiabatically) 
increases\footnote{This is what makes
the comparison safest
in the BPS case, due to the non-renormalization
properties of BPS state counting. In the non-BPS case, instead, one has to
compute the corrections to the microstate counting when
interactions are turned on.} from
$g_s=0$ to $g_s = g_c$.  

This entails coping with two main issues: i) computing the relevant
corrections to the tree-level results due to interactions,
ii) correctly defining and measuring the size of string states and
their {\em size distribution} and check its relation with the
horizon radius of the
expected correspondent black hole.

Refs.~\citen{stringblackhole, stringmassshifts} investigated these
issues within 
string theory. In fact, these problems where previously treated in 
approximated models, leading to a variety of
results \cite{selfgravstrngapprox}, but never in 
fully rigorous string formalism. 

In coping with ii), the traditional approach of defining a
size operator 
for the string failed to comply with the requirements of the theory. 
Ref.~\citen{stringblackhole}, instead, measured the size of
an object through a well-defined
operational procedure. The basic idea
is that,  
in a theory where the observables are scattering amplitudes,
the size of an object is measured by scattering probes\footnote{In
our case, a 
linear combination of graviton, dilaton and Kalb-Ramond field as
we are interested in the mass distribution of the object.} over
it. For mixed states (ensembles), the form factor
obtained in this way can be related to their size 
distribution.

In a microcanonical ensemble\footnote{Which appears to be the only
  well-defined one for closed string states due to the level matching
  conditions.}, the entropy of the superstring at zero coupling and
fixed (tree) mass 
level $M_0^2  =  Q_{L(R)}^2 +N_{L(R)}$, Neveu-Schwarz charges $Q_{L(R)}^i$ 
and size $R$ was found to be ($\alpha' = 4,\, \mathcal{N} =
\sqrt{N_L}+\sqrt{N_R} \nonumber$) 
 \begin{equation}
  S  =  \text{{\small$\pi \mathcal{N}\sqrt{d-1}-\frac{3\sqrt{d-1}}{8\mathcal{N} \pi}R^2 
   +\ln\left(\frac{R^{d-1}}{N_L^{\frac{d+2}{4}}N_R^{\frac{d+2}{4}}\mathcal{N}^{\frac{d}{2}}} \right)$}}
 \end{equation}

Turning on interactions ($g_s \neq 0$) yields corrections to
the free-string result. Neglecting
those related to $R$ (renormalization of the size), the relevant
ones come from the string mass-shift. 

The study of mass shifts in string theory
is very difficult because very few physical vertex operators
for massive string states are actually known (in covariant gauge) and
also because of the intrinsic complications in the
computations of  amplitudes \cite{oneloopold}. 
 
In order to
obtain the average squared 
mass shift for the relevant ensemble of states at fixed tree-level mass
charges and size, Ref.~\citen{stringmassshifts} exploited two key
properties of the string amplitudes.
First, the
unitarity of the string S-matrix and, then, the
modular and periodicity properties of the string amplitudes on the
torus, which allowed to determine a system of recurrent equations
for the amplitude itself\footnote{Namely for its coefficients, once
  the amplitude is expanded in powers of suitable variables, see
  Ref.~\citen{stringmassshifts}.}.  

Eventually, the average squared mass shift for the ensemble of string
states at fixed mass and charges\footnote{Again, we
  consider for simplicity
  $|Q^i_L|=|Q^i_R|=|Q^i|, \,\, N_L=N_R=N$. \label{Specifchargesmass}}
was found \cite{stringmassshifts} to be\footnote{This 
  formula tells us 
that the interaction in the effective Lagrangian will become important
-- of order one -- at $g_{se} \sim (M^2-Q^2)^{\frac{d-6}{8}}$ as
expected from field theory considerations. We are writing the result in
terms of the physical mass $M$, which is probably an even more
accurate estimate.} 
 \begin{equation}
  \overline{\Delta M^2}_{|_{N,Q}} = -g_s^2 (M^2-Q^2)^{1+\frac{3-D}{4}} \, ,
 \end{equation}
while the average squared mass shift for the ensemble of states at
fixed $M, Q^i, R$ is \cite{stringblackhole}
 \begin{equation}
  \overline{\Delta M^2}_{|_{M,Q,R}} = -g_s^2 (M^2-Q^2) R^{2-d} \, ,
 \end{equation}
absorbing an inessential proportionality constant in $g_s$.
We can then infer the corrections to the single-string
entropy $S$ when the $g_s$ is adiabatically increased \cite{stringblackhole}
 \begin{equation}
  e^{S} \sim   e^{2\pi\sqrt{d-1}\left(\sqrt{M^2-Q^2}+g_s^2 \frac{\sqrt{M^2-Q^2}}{2 R^{d-2}}-\frac{3}{32\pi^2\sqrt{M^2-Q^2}}R^2\right)}
 \end{equation}
In particular, we see that for 
 $
  2 ^{-1} g_s^2 \sqrt{M^2-Q^2}  R^{2-d} \geq 1 \, ,
 $
the entropy becomes dominated by strings whose size is
 $
  R \lesssim R_b \sim  
   (g_s^2 \sqrt{M^2-Q^2})^{\frac{1}{d-2}} \, ,
 $
which has the same perturbative formula of the horizon
radius\footnote{In the Einstein frame, which must be used to 
  compute the black hole entropy.} of the correspondent
black hole. It approaches the string length
at $g_s\sim (M^2-Q^2)^{-\frac{1}{4}}$, in accordance with the
correspondence principle.

The string entropy is dominated in particular by the states with the
minimum possible size. It can be shown \cite{stringblackhole} that
perturbation theory is sufficient to determine the minimal size only
in $d=3$, yielding
 $
  R_{\text{min}} = \frac{2}{g_s^2 (M^2-Q^2)} .
 $
It is finally interesting to observe that the relative average
one-loop mass shift 
for these minimum size states is of order 
$\frac{\overline{\Delta M^2}_{|_\text{{\tiny{$M,Q,R_{min}$}}}}}{M^2} \sim O(1)$,
possibly signaling a radical change in the relation between
entropy and mass for $g_s > g_c$ and hinting to a transition between a 
perturbative string regime and a black hole one. Finally, the corrected
formula for the entropy at leading order in $d=3$ is \cite{stringblackhole}
 \begin{equation}
  S(M, Q) \sim 2\pi\sqrt{2}\sqrt{M^2-Q^2}
      \left(1+\frac{g_s^4 (M^2-Q^2)}{4} \right) \, .
 \end{equation}

\section*{Acknowledgments}

I would like to thank very much Gabriele Veneziano and Paolo di
Vecchia for the many illuminating conversations about this subject of research.

\end{document}